# DESI data and refinement of standard recombination theory help solve the Hubble tension problem

A.V. Shepelev

Gubkin Russian State University of Oil and Gas, Moscow, Russia.

The values of the Hubble constant obtained from the data of the early and of the late Universe differ by 10%. Data obtained by DESI Collaboration can help solve this problem, establishing with high precision the value of the product of the Hubble constant h and comoving of sound horizon at the end of drag epoch to be equal to 101.8 Mpc. This value agrees very well with the earlier refinement of the standard recombination theory and the value of the Hubble constant $H_0$=73.5 km/s/Mpc obtained from measurements in the late Universe.

## Introduction

The Hubble tension has been an intriguing astrophysics problem in the last decade. According to measurements using CMB/BAO, the average value of the Hubble constant can be taken as $H_0 = 67.5 \, \text{km/s/Mpc}$. According to [1-4], its average value is approximately $H_0 = 73.5 \, \text{km/s/Mpc}$. It is fundamental that a higher value is obtained from the results of measurements in the late Universe, but a lower value is based on BAO and CMB, i.e. related to its early stages. The latest DESI data [5] provide an opportunity to clarify the question of the preferred value of the Hubble constant.

## Recombination rate and sound horizon

Interpretation of objective data based on ideas about the early Universe relies heavily on standard recombination theory (SRT)[1]. Its most important prediction is the rate of recombination and the resulting value of the sound horizon of the recombination epoch $r_*$ [6]. The latter is associated with the magnitude of comoving of sound horizon at the end of drag epoch $r_d$:

$$r_d \approx 1.018 r_*. \qquad (1)$$

The angular size of standard ruler $\theta$ depends on $r_d$:

$$\theta = \frac{r_d}{D}. \qquad (2)$$

Here, $D$ is a comoving distance from a present-day observer to the last scattering surface. It is the quantity $\theta = 0.0141$ that is the result of objective measurements.

According to SRT, the rate of recombination is slower than prescribed by the Saha law. SRT explains this by the slow dispersal of 2p-hydrogen state due to the high spectral energy density of resonance $Ly_\alpha$-radiation that occurs during recombination. $Ly_\alpha$-radiation, which originates during the transition of a hydrogen atom from the excited 2p state to the ground 1s state, is immediately absorbed by other atoms in the 1s state, transferring them to the 2p state. This leads to an overpopulation of atoms in 2p states, which are then easily ionized by equilibrium Planck background radiation. As a result, the rate of atom formation turns out to be low, since according to SRT the spectral density of energy $Ly_\alpha$-radiation is high owing to its slow escape solely due to the Hubble frequency shift. As a result, the recombination process is slower than predicted by the Saha law.

---

[1] When determining cosmological parameters based on the results of the Planck experiment [7], the authors explicitly note (quote): "Since the results of the Planck parameter analysis are crucially dependent on the accuracy of the recombination history, we have also checked, following Lewis et al. (2006), that there is no strong evidence for simple deviations from the assumed history. However, we note that any deviation from the assumed history could significantly shift parameters compared to the results presented here and we have not performed a detailed sensitivity analysis."

However, SRT does not take into account the process of excitation of the external degrees of freedom of neutral hydrogen during the interaction of $Ly_\alpha$-radiation with atoms. The fact is that in the field of isotropic background radiation, the atom is subjected to the action of a random force due to the transfer of momentum of the absorbed photon. The atom then returns to its original state, re-emitting a photon in a random direction. As a result, the average square of the atom's momentum increases if there is a dissipative force acting on the atom that satisfies the conditions of the fluctuation-dissipative theorem. The rate of change of the atom's momentum is proportional to the momentum $P$ (see calculations in [8]):

$$\frac{dP}{dt} = -P\frac{h\nu}{mc^2}B_{12}\rho, \tag{3}$$

where $B_{12}$ is the Einstein coefficient, $\rho$ is the spectral energy density of $Ly_\alpha$-radiation.

The fundamental fact is that the interaction of photons with atoms is discrete, since the average time between collisions of photons with an atom is much longer than the lifetime of an atom in an excited state. Therefore, the right side of the equation (3) presents the sum of the viscosity force and the random force. That is why the process cannot be described by a standard differential equation, but is described by the Langevin stochastic differential equation for the dependence of the average velocity of atoms on time [8,9]:

$$\frac{dV}{dt} = -\xi V + f(t). \tag{4}$$

Here, $\xi = \frac{g_2}{g_1}\frac{c}{8\pi m\nu^2}A\rho$, $A$ is the Einstein coefficient, $f(t)$ is a random function. The process can be considered as a Markov process. (It should be noted that the kinetic interaction of hydrogen atoms with radiation is similar to the interaction of a Brownian particle with liquid molecules.)

The Langevin equation solution gives the dependence of the average kinetic (thermal) energy of the hydrogen atom on time, on the radiation temperature $T_L$, and on the initial temperature of the atoms $T_P$, which can be assumed to be equal to the temperature of the Planck's background:

$$w(t) = \frac{3kT_L}{m} - \left(\frac{3kT_L}{m} - \frac{3kT_P}{m}\right)e^{-2\xi t}. \tag{5}$$

Accordingly, the energy of atoms per unit volume is

$$W(t) = \frac{3}{2}nk_B\Delta T\left(1 - e^{-\frac{t}{t_r}}\right). \tag{6}$$

Here,

$$t_r = \left(\frac{g_2}{g_1}\frac{cA\rho}{4\pi m\nu^2}\right)^{-1} \tag{7}$$

is a characteristic time [8].

Due to the requirement of energy balance, the kinetic energy obtained by atoms is related to the same decrease in radiation energy $\Delta W$, which manifests itself in a change in the radiation frequency:

$$\Delta W = \frac{3}{2}\frac{nk\Delta T}{t_r}\Delta t = \Delta(\Delta\nu\cdot\rho) = \frac{\Delta\nu}{\nu}\Delta\nu\cdot\rho. \tag{8}$$

Here, $\Delta\nu$ is the width of the Doppler band. This implies an estimate of the time during which, due to the exchange of photon pulses with atoms, the temperature $Ly_\alpha$-radiation decreases by half [8,10]:

$$\Delta t = \frac{128\pi}{3} \ln 2 \frac{g_1}{g_2} \frac{T_p}{\Delta T} \frac{\tau}{\lambda^3 n} . \tag{9}$$

Here, $\tau = 2.14 \cdot 10^{-9} s$ is the lifetime of the excited 2p-state, $\lambda = 0.1216 \cdot 10^{-6} m$ is the wavelength of $Ly_\alpha$-radiation.

Calculations carried out in [8.10] show that as a result of this process, $Ly_\alpha$-radiation exits the interaction with hydrogen atoms 2-3 orders of magnitude faster than due to the Hubble shift. The spectral density of $Ly_\alpha$-radiation decreases also. Therefore, its effect on the rate of recombination cannot be considered significant, and the recombination process corresponds quite accurately to the Saha law.

The higher the recombination rate, the greater the value of the redshift $z_*$ value corresponding to the recombination end, and the smaller the value of the acoustic horizon $r_*$:

$$r_* = \int_{z_*}^{\infty} \frac{c_s}{H_0 \sqrt{\Omega_r (1+z)^4 + \Omega_m (1+z)^3 + 1 - \Omega_r - \Omega_m}} dz . \tag{10}$$

$z_*$ is defined as the redshift corresponding to the maximum of the visibility function $g(z)$:

$$g(z) = \frac{d\tau(z)}{dz} \cdot \exp(-\tau(z)) . \tag{11}$$

Here,

$$\tau(z) = \int \frac{\sigma_T c n_e(z)}{(1+z) H(z)} dz \tag{12}$$

is the optical depth, $\sigma_T$ is the Thomson scattering crossection, $c_s$ is a speed of sound. An electron density $n_e(z)$ is determined by the Saha law.

**Conclusion**

Taking into account the excitation of the external degrees of freedom of neutral hydrogen gives a redshift value $z_{*1} = 1279$ for the moment of the end of recombination, determined according to Eqs. (10-12), see details in [11]. According to SRT, the value of $z_*$ is approximately 1090. Using Eqs. (1, 10), it is possible to determine the magnitude of the comoving of sound horizon at the end of drag epoch $r_{d1}$, because the value $r_d = 147.1$ Mpc predicted by SRT is known [5]:

$$r_{d1} = \frac{r_{*1}}{r_*} r_d \approx \frac{\sqrt{\frac{\Omega_m}{z_{*1}} + \Omega_r}}{\sqrt{\frac{\Omega_m}{z_*} + \Omega_r}} r_d = 0.936 r_d = 137.7 . \tag{13}$$

Here, according to the results of DESI, $\Omega_m = 0.295$ [5]; the value $\Omega_r$ is assumed to be $5 \cdot 10^{-5}$. The value of the Hubble constant $h$, obtained from measurements in the late Universe, is close to 0.735.

The latest published DESI results give the value of the product of the comoving of sound horizon at the end of drag epoch and the Hubble constant $h$ equal to 101.8 Mpc [5]. The comparison demonstrates a very good agreement between the previously obtained results [1-4, 10, 11] and the DESI data:

$$137.7 \cdot 0.735 = 101.2 \cong 101.8 .$$

This is an important argument confirming the value of the Hubble constant close to $H_0 = 73.5\,\text{km/s/Mpc}$.

**Acknowledgements**

The author is grateful to G.Kh. Kitaeva for the help in preparing this work.